\documentclass[a4paper,twocolumn,english,floatfix]{revtex4}
\usepackage{amsmath}
\usepackage{graphics}
\usepackage{graphicx}
\usepackage{color}

\begin{document}

\title{Cavity-assisted squeezing of a mechanical oscillator}

\date{\today}
\author{K. J\"ahne$^1$, C. Genes$^1$, K. Hammerer$^1$, M. Wallquist$^1$, E.S. Polzik$^2$ and P. Zoller$^1$}
\affiliation{$^1$ Institute for Theoretical Physics, University of Innsbruck, and Institute for Quantum Optics and Quantum Information,
Austrian Academy of Sciences, Technikerstrasse 25, 6020 Innsbruck, Austria\\
$^2$Niels Bohr Institute, QUANTOP, Danish Research Foundation Center for Quantum Optics, Blegdamsvej, DK-2100 Copenhagen, Denmark}

\begin{abstract}
We investigate the creation of squeezed states of a vibrating membrane or a movable mirror in an opto-mechanical system. An optical cavity is driven by squeezed light and
couples via radiation pressure to the membrane/mirror, effectively providing a squeezed heat-bath for the mechanical oscillator. Under the conditions of laser cooling to the ground state, we find an efficient transfer of squeezing with roughly 60\% of light squeezing conveyed to the membrane/mirror (on a dB scale). We determine the requirements on the carrier frequency and the bandwidth of squeezed light. Beyond the conditions of ground state cooling, we predict mechanical squashing to be observable in current systems.
\end{abstract}

\maketitle

\section{Introduction}

Recent progress in feedback and cavity-assisted cooling of micro-
and nano-mechanical resonators \cite{Metzger2004,Gigan2006,Arcizet2006b,Schliesser2008,Corbitt2007,Thompson2008,Teufel2008,Schliesser2009,Groeblacher2009}
shows that opto-mechanical systems are ultimately approaching the
quantum regime. Occupancy levels of around $30$ quanta of a motional
mode of a vibrating mirror have already been achieved experimentally
\cite{Groeblacher2009} via the sideband cooling technique and
limitations at the moment seem to be of a rather technical nature.
Similar to the case of a light field where the imprint of the
quantum regime is signaled by the generation of nonclassical
states such as squeezed states, generation of squeezing of a
nano-mechanical object can be a hallmark for quantum control of a
macroscopic, massive object \cite{Walls1983,Slusher1985,Wu1986}. On a more practical side,
nano-mechanical squeezing might have applications in ultrahigh
precision measurement experiments \cite{LaHaye2004} and detection of gravitational
waves \cite{Bradaschia1990}.

\begin{figure}[t]
\begin{center}
\includegraphics[width=0.68\columnwidth]{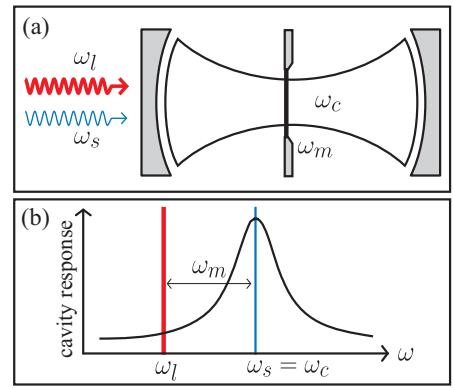}
\end{center}
\caption {(a) A mechanical mode of a dielectric membrane, oscillating at frequency $\omega_{\rm m}$, is coupled via radiation pressure to the cavity field of frequency $\omega_{\rm c}$. The cavity is driven by a laser of frequency $\omega_{\rm l}$  and, in addition, a much weaker squeezed vacuum field with central frequency $\omega_{\rm s}$.  (b) Optimal choice for the frequencies $\omega_{\rm l}$, $\omega_{\rm s}$, in order to transfer squeezing from the light field to the membrane motion: the laser frequency is chosen such that the cooling sideband is addressed, i.e. $\omega_{\rm l}=\omega_{\rm c}-\omega_{\rm m}$, whereas the center frequency of the squeezing has to equal the cavity resonance frequency $\omega_{\rm c}$.\label{SystemCavity}}
\end{figure}

Experimental squeezing of a nano-mechanical object was sofar only
achieved for a nonlinear Duffing resonator \cite{AlmogPRL98}. On the
theory side, squeezing of a linear nano-mechanical resonator was
proposed via coupling to an auxiliary nonlinear system, such as an
optical cavity containing an atomic medium \cite{IanPRA78}, a SQUID
loop \cite{ZhouPRL2006,ZhangArxiv2009} or a Cooper-pair box circuit
\cite{RablPRB70}. In principle squeezing could also be generated by teleportation of squeezed states of light \cite{Mancini2003} or atoms \cite{Hammerer2009} to the mechanical resonator. More direct approaches in opto-mechanical settings are based on modulated or parametric drive with or without feedback
\cite{ClerkNJP10,VitaliPRA65,WoolleyPRA78,TianNJP10}, or similarly,
modulated readout combined with a feedback loop \cite{RuskovPRB71}.

In this paper we analyze a scheme for generation of squeezing via
reservoir engineering in an opto-mechanical setup. Namely, starting
from a laser-cooled membrane, by accompanying the cooling beam with
a much weaker, squeezed vacuum input field, we show that the
membrane motion is driven into the steady-state of a squeezed
environment, which is a squeezed state. This method does not require
feedback or modulation of drive fields. It combines naturally with
requirements for ground state cooling, avoiding drive fields on the
blue side of the cavity resonance and associated issues concerning
dynamical stability. We show that, under the conditions of ground
state cooling, a significant transfer of squeezing from light to
mechanical degrees of freedom is possible. As a rule of thumb, our studies predict a
transfer of approximately 60\% of squeezing on a logarithmic scale,
e.g. 6\,dB of light squeezing would result in 4\,dB of mechanical
squeezing. Optimal transfer is achieved when the central frequency of
squeezing is resonant with the cavity mode. This resonance condition has
to be fulfilled within a tolerance on the order of the spectrally
broadened linewidth of the mechanical oscillator, which is
typically smaller than the cavity linewidth. We take full account
of a finite bandwidth of squeezed light, which makes the dynamics
essentially non-Markovian. We find that an optimal finite bandwidth
exists if the sidebands are only poorly resolved, while there is
practically no dependence on bandwidth in the resolved sideband
limit. Finally, we show that beyond the regime of ground state
cooling squashed states \cite{Wiseman1999} of the mechanical
resonator could be observed under present experimental conditions.

The transfer of quantum noise properties from the radiation field to the mechanical degrees of freedom lies also at the heart of the schemes presented in \cite{PinardEPL72,ZhangPRA68}, which aim for the creation of entanglement between two movable mirrors. With experiments entering the quantum regime of cooling close to the ground state, we see the present work as a natural first step towards such more demanding protocols. We provide here for the first time a careful analysis of the effects of mismatch in resonance conditions and finite squeezing bandwidth.

The paper is structured as follows: In Section II we derive the exact
equations of motion (quantum Langevin equations) of the coupled light-membrane system.
In Section III we derive the optimal conditions for squeezing transfer and
find simple analytical expressions for the membrane variances in an adiabatic limit where the cavity field is eliminated and under the assumption of resolved-sideband limit. We compare the results with an exact numerical
solution and analyze the domain of validity of the adiabatic elimination of the field.
Conclusions are presented in Section IV and analytical results outside  the resolved sideband limit are listed in the Appendix.

\section{Opto-mechanical system driven by squeezed light\label{Sec:System}}

We consider the dispersive opto-mechanical system illustrated in
Fig. \ref{SystemCavity}, where a vibrating dielectric membrane is
placed between the two fixed end-mirrors of a laser-driven, single sided
Fabry-Perot cavity \cite{Thompson2008}. The theory presented is in
principle also applicable to a Fabry-Perot cavity with a light,
movable end-mirror \cite{Genes2008}, or to a vibrational mode of a
whispering gallery mode cavity \cite{Kippenberg2005,Schliesser2006}.
The opto-mechanical coupling strength in general depends on the
particular geometrical factors of the system under consideration
while a prototypical system can always be modeled by a simple
Hamiltonian
\begin{multline}
H=\hbar\omega_{\rm c} c^\dag c+\hbar\omega_{\rm m0}b^\dag b-\hbar
g c^\dag
c(b^\dag+b)\\
+i\hbar \left(Ee^{-i\omega_{\rm l} t}c^\dag-E^*e^{i\omega_{\rm l}t}c\right)
\label{Eq:Hamiltonian}.
\end{multline}
The first term in the Hamiltonian gives the energy of a single
cavity mode at an optical frequency $\omega_{\rm c}$,
characterized by the annihilation operator $c $ satisfying the
commutation relation $[c,c^\dag]=1$. The second term gives the
energy of the motional mode of the membrane, at a resonance
frequency $\omega_{\rm m0}$,
where $b$ is the annihilation operator for vibrational quanta. The
third term describes the membrane-light radiation pressure coupling
with strength $g=2R (\bar{x}_{\rm m}/L)\omega_{\rm c}$, where
$\bar{x}_{\rm m}=\sqrt{\hbar/2 m\omega_{\rm m0}}$ is the zero-point
motion of a membrane mode of effective mass $m$. The geometrical
factors $L$ and $R$ are the effective cavity length and reflectivity
of the membrane. In our setup the membrane is placed at the point of
maximal linear coupling to the cavity field \cite{Thompson2008}.
The last term shows driving of the cavity field with a laser at
frequency $\omega_{\rm l}$ and strength $E$ related to the laser
power $P$ by $|E|=\sqrt{2P\kappa/\hbar\omega_c}$, where $\kappa$ is
the cavity amplitude decay rate.

In addition to the evolution described by the Hamiltonian in
Eq.~(\ref{Eq:Hamiltonian}) the system is also subjected to random noise forces due to fluctuations of the phononic heat bath of the membrane and to quantum fluctuations of the radiation field. The main idea here is to deliberately shape the noise properties of the latter bath by driving the cavity with a
squeezed vacuum field in parallel to the coherent field. The effect of squeezed vacuum noise on the cavity field is
included in a non-Markovian noise model with damping rate $\kappa$ and
input noise operator $c_{\rm in}(t)$ whose two-time correlation functions are
\begin{equation}\label{ainainPhys}
\begin{split}
\langle\bar{c}_{\rm in}(t+\tau)\bar{c}_{\rm in}(t)\rangle&=\frac{M}{2}\frac{b_x b_y}{b_x^2+b_y^2}\left(b_y e^{-b_x|\tau|}+b_x e^{-b_y |\tau|}\right)\\
\langle \bar{c}^\dag_{\rm in}(t+\tau)\bar{c}_{\rm in}(t)\rangle&=\frac{N}{2}\frac{b_x b_y}{b_y^2-b_x^2}\left(b_y e^{-b_x|\tau|}-b_x e^{-b_y |\tau|}\right).
\end{split}
\end{equation}
The noise operators $\bar{c}_{\rm in}(t)=e^{i\omega_{\rm s}t} c_{\rm in}(t)$ refer
to a frame rotating at the carrier frequency of squeezing
$\omega_{\rm s}$ and satisfy the canonical commutation
relation $[ \bar{c}_{\rm in}(t),\bar{c}^\dag_{\rm
in}(t')]=\delta(t-t')$ \cite{Ritsch1988}. Parameters $N$ and $M$
determine the degree of squeezing, while $b_x$ and $b_y$ define the squeezing bandwidth. The connection of these parameters to the properties of the optical parametric oscillator cavity generating the squeezing are summarized in Appendix \ref{App:inputfield}. In particular for pure squeezing there are only two independent parameters, as in this case $|M|^2=N(N+1)$ and $b_y=b_x\sqrt{2\left(N+|M|\right)+1}$.
While we will derive all results for a finite bandwidth, it will be
instructive and more transparent to consider in certain cases the
white noise limit for squeezing, that is $b_x,b_y\to\infty$ while
keeping $N,M$ constant. In this case the correlation functions of
Eq. (\ref{ainainPhys}) are simplified to $\langle \bar{c}_{\rm
in}(t+\tau)\bar{c}_{\rm in}(t)\rangle\to M \delta(\tau)$ and
$\langle \bar{c}^\dag_{\rm in}(t+\tau)\bar{c}_{\rm in}(t)\rangle\to
N \delta(\tau)$.

The thermal bath affecting the motion of the membrane is Brownian
and consequently non-Markovian \cite{Brownian}. However, in the high
temperature limit ($2kT\gg \hbar\omega_{\rm m0}$), which is
applicable to this system even for cryogenic temperatures,
the noise can be described using a Markovian model with membrane
loss rate $\gamma_{\rm m}$. The noise operators $b_{\rm in}(t)$ are
characterized by the correlation functions
\begin{eqnarray}
\langle b^\dag_{\rm in}(t)b_{\rm in}(t')\rangle&=&n_{\rm th}\delta(t-t')\label{Eq:ThermalNoise}
\end{eqnarray}
and commutation relations $[ b_{\rm in}(t),b^\dag_{\rm
in}(t')]=\delta(t-t')$. Here we denoted by $n_{\rm th}$ the average
thermal occupation number of the mechanical mode.

Having specified the Hamiltonian and noises affecting the
membrane-light system, we proceed now to analyze its dynamics using
the quantum Langevin equations formalism
\begin{eqnarray}\label{Eq:NonlinearEOM}
\dot{c}&=&-\left(i\left[\omega_{\rm c}-g(b^\dag+b)\right]+\kappa\right)c
+Ee^{-i\omega_{\rm l}t}+\sqrt{2\kappa}c_{\rm in}(t)\nonumber\\
\dot{b}&=&-(i \omega_{\rm m0}+\gamma_{\rm m})b +igc^\dag
c+\sqrt{2\gamma_{\rm m}}b_{\rm in}(t).
\end{eqnarray}
In the following we perform a transformation to a frame rotating at the laser frequency
$\omega_{\rm l}$, so that $\tilde{c}(t)= c(t)e^{i\omega_{\rm l}t}$.
We linearize operators around the steady state values
$\tilde{c}=\langle\tilde{c}\rangle_{\rm ss}+\delta \tilde{c}$,
$b=\langle b\rangle_{\rm ss}+\delta b$ such that the fluctuations
$\delta b$ and $\delta \tilde{c}$ have zero mean.
We find the steady state value for the cavity amplitude $\langle
\tilde{c}\rangle_{\rm ss}= E/(\kappa+i\Delta)$, where the effective
detuning is
\[\Delta=\omega_{\rm c}-\omega_{\rm l}-\frac{2g^2\omega_{\rm m0}}{\omega_{\rm m0}^2+\gamma_{\rm m}^2}\langle \tilde{c}^\dag \tilde{c}\rangle_{\rm ss}.\]
For simplicity we take $\langle \tilde{c}\rangle_{\rm ss}$ to be
real and positive; this can be achieved by an appropriate choice of
the laser phase.
Similarly, we find for the mechanical oscillator $\langle
b\rangle_{\rm ss}= \left(g/(\omega_{\rm m0}-i\gamma_{\rm
m})\right)\langle \tilde{c}^\dag \tilde{c}\rangle_{\rm ss}$. Let us
remark, that the intra-cavity occupation at the steady state,
$\langle \tilde{c}^\dag\tilde{c}\rangle_{\rm ss}$, contains two
contributions, one from the laser drive and the other from inserting
squeezed light.
The laser contributes a number of $|\langle\tilde{c}\rangle_{\rm ss}|^2$ photons, which -- in
present setups -- are typically more than $10^5$ photons.
One can show, that the contribution from squeezed light is given by
$N\left(b_xb_y/(b_y^2-b_x^2)\right)\left[b_y/(b_x+\kappa)-b_x/(b_y+\kappa)\right]\leq
N$. Squeezed vacuum exhibiting a large noise reduction of -10 dB below shot noise level \cite{Vahlbruch} thus contributes at most $N\sim 2$ intra-cavity photons. The contribution of the squeezed vacuum to the intra-cavity photon number is thus negligible.

We can proceed now to linearize the equations of motion
\eqref{Eq:NonlinearEOM} following the standard treatments of
opto-mechanical coupling \cite{Genes2008}. A mean field expansion
around the large coherent intracavity amplitude $\langle
\tilde{c}\rangle_{\rm ss}$ yields a linear set of equations
\begin{eqnarray}
\delta\dot{\tilde{c}}&=&-\left(i\Delta+\kappa\right)\delta \tilde{c}
+\frac{iG}{2}\left(\delta b^\dag+\delta b\right)
+\sqrt{2\kappa}\bar{c}_{\rm in}(t)e^{i\Delta_{\rm s}t}\nonumber\\
\delta\dot{b}&=&-(i \omega_{\rm m0}+\gamma_{\rm m})\delta b
+\frac{iG}{2} \left(\delta\tilde{c}^\dag +\delta \tilde{c}\right)
+\sqrt{2\gamma_{\rm m}}b_{\rm in}(t)\nonumber\\
\label{Eq:Fluctuations}
\end{eqnarray}
where we introduced the detuning between the laser and the center frequency of the squeezing
\[\Delta_{\rm s}=\omega_{\rm l}-\omega_{\rm s}\]
and the effective opto-mechanical coupling
\[G=2g\langle \tilde{c}\rangle_{\rm ss}.\]

Finally we move to a frame rotating at the detuning $\Delta$ for
cavity operators and rotating at a {\it shifted} mechanical
frequency $\omega_{\rm m}=\omega_{\rm m0}-\Omega$ for the membrane,
i.e. we transform $\delta c_{\rm I}(t)= \delta
\tilde{c}(t)e^{i\Delta t}$ and $\delta b_{\rm I}(t)=\delta b(t)
e^{i\omega_{\rm m}t}$. We use the shifted frequency $\omega_{\rm m}$
to accommodate for the optical spring effect \cite{Genes2008}, which
introduces a shift $\Omega$ of the bare resonance frequency
$\omega_{\rm m0}$, as we will see later on. Typically this shift is
negligible for oscillators at the level of MHz, but it turns out to
play an important role in the condition of optimal squeezing
transfer. The exact expression of $\Omega$ will be detailed in the
next Section. The quantum Langevin equations that form the starting
point of our analysis can now be written
\begin{eqnarray}
\delta\dot{c}_{\rm I}&=&-\kappa\delta c_{\rm I}+\frac{iG}{2}e^{i\Delta t}\left(\delta b_{\rm I}^\dag e^{i\omega_{\rm m}t}+\delta b_{\rm I}e^{-i\omega_{\rm m}t}\right)\nonumber\\
&&+\sqrt{2\kappa}\bar{c}_{\rm in}(t)e^{i(\Delta+\Delta_{\rm s})t}\label{QLEcItilde}\\
\delta\dot{b}_{\rm I}&=&-\left(\gamma_{\rm m}+i\Omega\right)\delta b_{\rm I}
+\frac{iG}{2}e^{i\omega_{\rm m}t}
\left(\delta c_{\rm I}^\dag e^{i\Delta t}
+\delta c_{\rm I}e^{-i\Delta t}\right)\nonumber\\
&&+\sqrt{2\gamma_{\rm m}}b_{\rm in,I}(t)\label{QLEbI}.
\end{eqnarray}

\section{Mechanical Squeezing\label{Sec:AnalyticSqueezing}}

The set of Eqs. (\ref{QLEcItilde}, \ref{QLEbI}) can, in principle,
be solved exactly. However, the resulting expressions are rather
cumbersome and do not offer the necessary physical insight into the
transfer of squeezing from the light to the motion of the membrane.
Consequently we derive an approximate solution in the
perturbative limit of weak opto-mechanical coupling $G\ll\kappa$, which we then compare with the exact
solution of Eqs.~(\ref{QLEcItilde},\ref{QLEbI}) for particular sets of parameters.

As a first main step in our analysis, we note that squeezing one
of the membrane's variances below the shot noise limit can only be expected under conditions where the state of the membrane is already close to the ground state. Therefore, we require simultaneous
laser cooling of the motion which can be achieved with the condition
\[\Delta =\omega_{\rm m},\]
i.e., in the sideband cooling regime \cite{Marquardt2007,Wilson-Rae2007,Genes2008}.

Under the assumption of fast cavity dynamics on the time scale of the cavity-membrane coupling, $G/\kappa \ll 1$, we adiabatically eliminate the cavity mode (see also \cite{Wilson-Rae2007}).
Formally solving for the cavity dynamics (\ref{QLEcItilde}), and keeping terms up to $O(G/\kappa)$ when evaluating the integrals over the membrane motion, we find an approximate solution for the cavity mode $\delta c_{\rm I}(t)$,
\begin{multline}
\delta c_{\rm I}(t)=\frac{iG}{2\kappa}\delta b_{\rm I}(t)
+\frac{i G}{2(\kappa+2i\omega_{\rm m})}e^{2i\omega_{\rm m}t}\delta b_{\rm I}^\dag(t)\\
+\sqrt{2\kappa}\int\limits_0^t d\tau e^{-\kappa(t-\tau)} e^{i(\omega_{\rm m}+\Delta_{\rm s})\tau}\bar{c}_{\rm in}(\tau).
\label{Eq:SolC}
\end{multline}
Inserting the solution Eq.~(\ref{Eq:SolC}) into the membrane equation of motion, Eq.~(\ref{QLEbI}),
the result is an effective equation of motion for the membrane
\begin{align}
\delta \dot{b}_{\rm I}(t)&=-\gamma_{\rm eff}\delta b_{\rm I}(t)
+\left(\Gamma+i\Omega\right)e^{2i\omega_{\rm m}t}\delta b_{\rm I}^\dag(t)
+\sqrt{2\gamma_{\rm m}}b_{\rm in,I}(t)\nonumber\\
&\quad\quad
+\frac{ iG\sqrt{\kappa}}{\sqrt{2}}\int\limits_0^t d\tau e^{-\kappa(t-\tau)}
\left[e^{i(\omega_{\rm m}+\Delta_{\rm s})\tau}\bar{c}_{\rm in}(\tau)\right.\nonumber\\
&\quad \label{Eq:QLEAdEl}
\hspace{2.3cm}\left.+e^{2i\omega_{\rm m}t}e^{-i(\omega_{\rm m}+\Delta_{\rm s})\tau}\bar{c}^\dag_{\rm in}(\tau)\right].
\end{align}

The first term on the right hand side describes damping at an effective decay rate,
\[\gamma_{\rm eff}=\gamma_{\rm m}+\Gamma,\]
which is the sum of the intrinsic decay rate $\gamma_{\rm m}$ of the membrane due to its coupling to the thermal bath, and the radiation pressure induced decay rate $\Gamma$, defined by
\[\Gamma=\frac{G^2}{4\kappa}\frac{4\omega_{\rm m}^2}{\kappa^2+4\omega_{\rm m}^2}.\]
The enhanced decay rate $\gamma_{\rm eff}$ is the signature of
sideband cooling. For high $Q$ mechanical oscillators and efficient
laser cooling we have $\Gamma\gg\gamma_{\rm m}$ such that we will
take $\gamma_{\rm eff}\simeq\Gamma$ in the following. The adiabatic
elimination of the cavity mode also gives rise to a shift $\Omega$
of the membrane frequency, which is given by
$\Omega=\Gamma(2\kappa/\omega_{\rm m})$. The second term on the
right hand side of Eq.~\eqref{Eq:QLEAdEl} is a counter-rotating
term; it makes a negligible contribution as
$\Omega,\Gamma\ll\omega_{\rm m}$ under the present assumptions,
consequently we can neglect it in the following.

The last term in Eq. (\ref{Eq:QLEAdEl}) corresponds to the desired
squeezed, non-Markovian Langevin force driving the motion of the
membrane. We are interested in an optimal transfer of the squeezing
properties to the mechanical system. An important question is, how
the center frequency of the squeezing $\omega_{\rm s}$ is to be
chosen with respect to the cavity frequency. In order to get an
efficient coupling of $b_{\rm I}$ to the squeezed noise force
$\bar{c}_{\rm in}$, the time dependence of the integrand in the last
term of Eq. (\ref{Eq:QLEAdEl}) suggests the choice
\[\Delta_{\rm s} =-\omega_{\rm m},\]
which amounts to a squeezing spectrum centered at the cavity
resonance frequency $\omega_{\rm s}=\omega_{\rm c}$. The coupling to
$\bar{c}^\dag_{\rm in}$ will then be fast oscillating and make a
negligible contribution. Below we will give a more physical picture for this resonance condition along with a discussion of the tolerance to a mismatch in this condition.

In the following we restrict our discussion to the resolved sideband
limit case, i.e. when the resolved sideband parameter is
$\eta=\kappa/\omega_{\rm m}\ll1$. In this regime, cooling of the
membrane is optimal and efficient squeezing is expected.
For simplicity of presentation, we keep only the zeroth order
terms in $\eta$, which results in a simple form for the expression
of the optical cooling rate $\Gamma=G^2/(4\kappa)$. The full
analytical expressions valid beyond the resolved sideband limit are
presented in Appendix \ref{Sec:AppGeneralCorrelations}.
The solution of Eq.~(\ref{Eq:QLEAdEl}) in the long time limit is
\begin{eqnarray}
\delta b_{\rm I}(t)&=&i \delta b_{\rm I,sq}(t)+\sqrt{\frac{\gamma_{\rm m}}{\gamma_{\rm eff}}}\delta b_{\rm I,th}(t)\label{Eq:Solb}
\end{eqnarray}
and exhibits a squeezing and a thermal contribution. The contribution due to the radiation pressure coupling to the squeezed environment is
\begin{multline}
\delta b_{\rm I,sq}(t)=\sqrt{2\gamma_{\rm eff}}\kappa
\int\limits_0^t d\tau e^{-\gamma_{\rm eff}(t-\tau)}
\int\limits_0^\tau d\tau'
e^{-\kappa(\tau-\tau')}\cdot\\
\times\left[e^{2i\omega_{\rm m0}\tau}\bar{c}^\dag_{\rm
in}(\tau')+\bar{c}_{\rm in}(\tau')\right]\label{Eq:Solbsq}
\end{multline}
and the thermal contribution is
\begin{eqnarray}
\delta b_{\rm I,th}(t)&=&\sqrt{2\gamma_{\rm eff}}
\int\limits_0^t d\tau e^{-\gamma_{\rm eff}(t-\tau)}
b_{\rm in,I}(\tau)\label{Eq:Solbtherm}.
\end{eqnarray}
The two noise processes are uncorrelated such that their effects to the steady state statistics will simply add up.

\begin{figure}[t]
\begin{center}
\includegraphics[width=1\columnwidth]{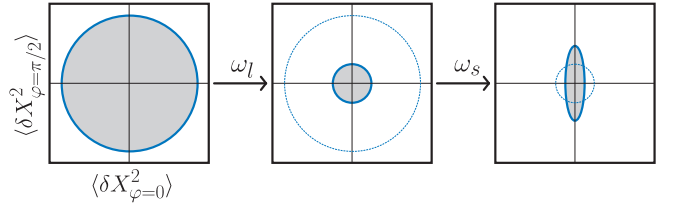}
\end{center}
\caption{Schematic phase-space picture of mechanical squeezing via reservoir engineering with squeezed light: A mechanical resonator in thermal equilibrium with equal variance in conjugate quadratures $\delta X_{\varphi=0}$ and $\delta X_{\varphi=\pi/2}$ (left) is laser sideband cooled by a coherent driving field at frequency $\omega_l$ reduces the mechanical variances $\langle\delta X_{\varphi}\delta X_{\varphi}\rangle$ close to the ground state variances (center). A second, squeezed vacuum field at frequency $\omega_s$ drives the system into a state with (anti)squeezed variances in conjuagte variables (right).}\label{Fig:Squeezing}
\end{figure}

\begin{figure*}[t]
\begin{center}
\includegraphics[width=0.65\textwidth]{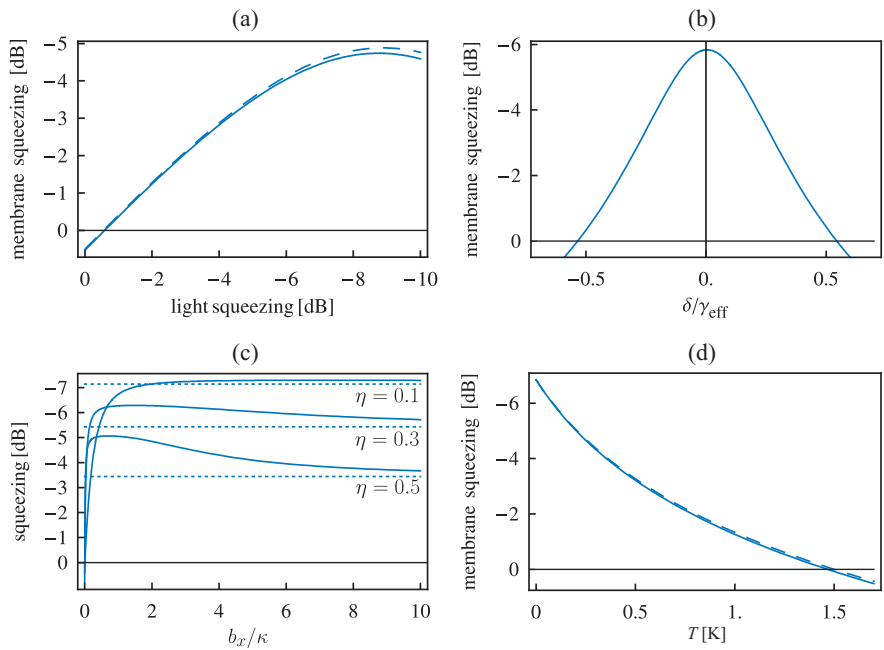}
\end{center}
\caption{(a) Efficiency of squeezing transfer from light to the membrane (in dB) for the following set of parameters $Q=10^7$, $m=1$\,ng, $\omega_{\rm m0}=2\pi\times1$\,MHz, $\kappa=2\pi\times380$\,kHz, $G=2\pi\times110$\,kHz and $T=100$\,mK. The analytical result of Eq.~(\ref{Eq:XXCorrRS}) (dashed line) agrees well with the exact numerical result (full line). (b) Squeezing of the membrane vs. deviations from the resonance, $\delta=\Delta_{\rm s} +\omega_{\rm m}$. Mechanical squeezing requires the resonance condition $\delta=0$ to be fulfilled within the effective mechanical linewidth $\gamma_{eff}$. (c) Squeezing transfer vs. bandwidth of squeezing $b_x$ for various values of the sideband parameter $\eta=\kappa/\omega_{\rm m}$ (full lines). Dashed lines give the corresponding value for white squeezed input noise. Outside the resolved sideband regime the bandwidth plays a role in optimizing the transfer of squeezing. (d) Dependence of the generated motional squeezing on the membrane environment temperature.} \label{Fig:Plots}
\end{figure*}

In order to see the effect of squeezing we need to evaluate the variances of the generalized quadrature operator
\begin{eqnarray}
\delta X_{\varphi,{\rm I}}(t)&=&\frac{1}{\sqrt{2}}\left(e^{i\varphi}\delta b_{\rm I}(t)+e^{-i\varphi}\delta b_{\rm I}^\dag(t)\right)\label{Eq:GenQuadrature},
\end{eqnarray}
which specializes for $\varphi=0$ to the usual position operator
$\delta q_{\rm I}(t)$ and for $\varphi=-\pi/2$ to the momentum
operator $\delta p_{\rm I}(t)$, both taken in a rotating frame at frequency $\omega_{\rm m}$. In Appendix
\ref{Sec:AppGeneralCorrelations} we evaluate the quadrature
correlations for finite bandwidth squeezing, that is when $\delta
b_{\rm I,sq}(t)$ is a non-Markovian noise process. The result is
very intuitive in the limit of squeezed white noise where the
quadrature correlations take a simple form
\begin{multline}
\langle\delta X_{\varphi,{\rm I}}(t)\delta X_{\varphi,{\rm
I}}(t)\rangle=
\left(N+\frac{1}{2}-{\rm Re}\left\{M e^{2i\varphi}\right\}\right)\\
+\frac{\gamma_{\rm m}}{\gamma_{\rm eff}}\left(n_{\rm th}
+\frac{1}{2}\right)\label{Eq:XXCorrRS}.
\end{multline}
The interpretation of Eq. (\ref{Eq:XXCorrRS}) is quite straightforward. In the first term we recognize the squeezing properties of the input field
\[
\langle x_{\varphi,{\rm in}}(t)x_{\varphi,{\rm
in}}(t-\tau)\rangle=\left(N+\frac{1}{2}+{\rm Re}\{Me^{2i\varphi}\}\right)\delta(\tau),
\]
where the light quadratures are $x_{\varphi,{\rm
in}}(t)=\frac{1}{\sqrt{2}}\left(e^{i\varphi}\bar{c}_{\rm
in}(t)+e^{-i\varphi}\bar{c}^\dag_{\rm in}(t)\right)$ and we used Eq.
(\ref{ainainPhys}) in the white noise limit. The second term is the
thermal variance, suppressed by a factor $\gamma_{\rm m}/\gamma_{\rm
eff}$, as is familiar from the opto-mechanical laser sideband
cooling \cite{Wilson-Rae2007,Marquardt2007,Genes2008}. In fact, for
the particular case of no squeezing ($M=N=0$), the final occupation
number $n_{\rm f}=(1/2)(\langle\delta q_{\rm
I}(t)^2\rangle+\langle\delta p_{\rm I}(t)^2\rangle-1)$ is given by
$n_{\rm f}\simeq \kappa^2/(2\omega_{\rm m0})^2\ll 1$, which is the
well-known residual occupancy for resolved sideband cooling as
obtained in Refs. \cite{Wilson-Rae2007,Marquardt2007,Genes2008}.
Figure ~\ref{Fig:Squeezing} provides a phase space illustration of
our result.

The simple result Eq. (\ref{Eq:XXCorrRS}) is valid only in the limit of squeezed white noise and to zeroth order in the sideband parameter $\eta$. For a realistic discussion of the main obstacles to obtain
perfect squeezing, we extend our treatment beyond the resolved-sideband regime. Our first observation is that for a nonzero $\eta$, even for a large cooling efficiency (where we can still neglect the contribution of the phononic heat bath), the resulting squeezed state is thermal. We find that $N$ of the pure squeezed state in \eqref{Eq:XXCorrRS} is replaced by $N'=N[1+\eta/(\eta+4)]+\eta/(2[\eta+4])$ while $M$ is unchanged. For the resulting squeezed mechanical state we thus have $|M|^2\leq N'(N'+1)$, which is the signature of a mixed, squeezed state. Thus, for a given finite sideband parameter $\eta$, we expect that for very large squeezing of light, mechanical squeezing eventually starts to degrade due to an increasing impurity.

In Fig.~\ref{Fig:Plots} we compare the analytical results for arbitrary bandwidth squeezing, given by Eq.~(\ref{Eq:XXCorrGen}) and squeezed white noise, given by Eq.~(\ref{Eq:XXCorrWhite}) with
numerical results obtained by exactly solving the equation of motion Eqs.~(\ref{QLEcItilde}, \ref{QLEbI}) in steady state. In Fig. \ref{Fig:Plots}a, we study the squeezing transferred to the membrane as
a function of the squeezing of the input light
(in dB). The results obtained via the exact numerical
method are compared with the approximate results of Eq. (\ref{Eq:XXCorrRS}). Under the conditions of ground state cooling, we see an efficient transfer of squeezing, with about 60\% of light squeezing being transferred to the mechanical system. As expected for the reasons given above, for a high amount of light squeezing the transfer degrades again.

Of great importance is the sensitivity
of the transfer to deviations from the resonance conditions. It is
well known \cite{GenesSim08} that cavity cooling is fairly robust to
deviations from the optimal cooling condition $\Delta = \omega_{\rm
m}$. Here we find, as shown in Fig. \ref{Fig:Plots}b where the
membrane squeezing is plotted vs. the detuning $\delta=\Delta_{\rm
s} +\omega_{\rm m}$, that fairly small deviations (of order
$\gamma_{\rm eff}$) from the condition of resonant squeezing
transfer lead to washed out membrane squeezing. This result can be understood from the fact that $\gamma_{\rm eff}$ gives the bandwidth of the mechanically scattered cooling sideband, such that a
deviation $\delta>\gamma_{\rm eff}$ would mean that the central
frequency of the squeezing input completely misses the sideband
cooling spectrum.

The results
shown in Fig. \ref{Fig:Plots}c indicate that in general there is an optimal
squeezing bandwidth for which the transfer from light to membrane is
maximized, but for small $\eta$ the finite bandwidth result of Eq.
(\ref{Eq:XXCorrGen}) (full lines) does not differ much from the infinite
bandwidth limit result (dashed lines).
For a large bandwidth which fully covers the motional sidebands, $b_x \gg
\omega_{\rm m}$, the membrane sees only white squeezed input noise, whereas for
smaller bandwidth, the crucial question is whether the squeezed input will
touch the heating sideband or not. For small $\eta$, as we see in Fig.
\ref{Fig:Plots}c, the width is not a big issue, since the heating sideband is
anyway weak, whereas for large $\eta$ the squeezing transfer is much improved
for an optimal, finite bandwidth where the strong heating sideband is avoided. For the
same reason, the ratio of the optimal bandwidth to $\kappa$ decreases for large
$\eta$.

Finally, we investigated the squeezing transfer as a function of the
environmental temperature. As shown in Fig. \ref{Fig:Plots}d it is
of great importance to provide a cold environment where ground-state
cooling of the membrane is allowed. However, as ground-state cooling
of membranes has yet to be demonstrated, we investigated the
squashing \cite{Wiseman1999} of the membrane state for existing
experimental parameters. In the context of existing cavity-assisted
mirror/membrane cooling experiments an occupancy of around $n_{\rm
f}=10$ is in sight. This can be achieved for the set of parameters
$Q=10^7$, $m=10$\,ng, $\omega_{\rm m0}=2\pi\times1$\,MHz,
$\kappa=2\pi\times125$\,kHz, $G=2\pi\times21$\,kHz and $T=4$\,K, and
corresponding cavity finesse of $6\times10^4$, cavity length
$L=20$\,mm and a circulating power of $P=3$\,mW. For a squashed
membrane state transferred from 6\,dB of squeezed vacuum input
light, the imbalance in variances is about 20\%. This imbalance will
be sensitive to the phase of the squeezed vacuum input and should be
easily detectable for the present shot-noise limited detection
efficiencies \cite{Schliesser2008b}. The basic principle of state engineering of a mechanical oscillator via reservoir engineering of quantum fluctuations of the radiation field, should thus be readily observable.

\section{Conclusion}

In conclusion, we have analyzed the conditions under which transfer
of squeezing from light to a membrane within an optical cavity is
possible and optimal. We have found that the best operating regime
is in the typical resolved-sideband limit for opto-mechanical
cooling, and that optimal transfer occurs when the squeezed light is
fed into the cavity at resonance. We have also shown that the
squeezed light bandwidth is a factor in the efficiency of the
transfer when one operates outside the strictly resolved-sideband
limit. However, we conclude that under ideal cooling conditions, as
long as the squeezing bandwidth exceeds the laser-induced cooling
rate of the membrane, the squeezed white noise description is
perfectly valid.

We thank M. Aspelmeyer for discussions. Support by the Austrian Science Fund through SFB FOQUS, by the Institute for Quantum Optics and Quantum Information, by the European Union through project EuroSQIP and by FET-Open grant HIDEAS (FP7-ICT-221906) is acknowledged.


\appendix

\section{Parametrization of squeezed light\label{App:inputfield}}

A squeezed light field described by the correlation functions (\ref{ainainPhys}) is routinely produced in an optical parametric oscillator (OPO) driven below threshold \cite{Collett1984}. The parameters $M,N$ and bandwidths $b_x,b_y$ of the squeezed light field (\ref{ainainPhys}) are related to the susceptibility $\epsilon$ of the OPO and the damping rate $\gamma$ of the OPO cavity by $b_x=\gamma/2-|\epsilon|$ and $b_y=\gamma/2+|\epsilon|$, and by $M=\left(\epsilon\gamma/2\right)\left[1/b_x^2+1/b_y^2\right]$ and
$N=\left(|\epsilon|\gamma/2\right)\left[1/b_x^2-1/b_y^2\right]$.
From this definition it is clear that $N\geq 0$; furthermore, the stability of the OPO cavity requires $b_x\geq 0$. The chosen parametrization satisfies the relation for a pure squeezed state, $|M|^2=N(N+1)$, which is the maximum squeezing limit of the general property $|M|^2\le N(N+1)$ for squeezed noise. Note that two of the parameters are redundant for the given case, as discussed in Section \ref{Sec:System}.

\section{Beyond the resolved sideband limit\label{Sec:AppGeneralCorrelations}}

Here we present the analytical results of Section \ref{Sec:AnalyticSqueezing} generalized to arbitrary ratios  of $\eta$. The adiabatic elimination of the cavity mode leaves us with the effective equation of motion (\ref{Eq:QLEAdEl}) for the membrane mode and its solution (\ref{Eq:Solb}) which here generalizes to,
\begin{eqnarray}
\delta b_{\rm I}(t)&=&\frac{iG}{\sqrt{\gamma_{\rm eff}\kappa}}\frac{\omega_{\rm m}}{\sqrt{\kappa^2+4\omega_{\rm m}^2}}\delta b_{\rm I,sq}(t)+\sqrt{\frac{\gamma_{\rm m}}{\gamma_{\rm eff}}}\delta b_{\rm I,th}(t).\nonumber
\end{eqnarray}
The contribution due to the squeezed environment is now given by (cf. Eq. (\ref{Eq:Solbsq})),
\begin{eqnarray}
\delta b_{\rm I,sq}(t)&=&\sqrt{2\gamma_{\rm eff}}\kappa\sqrt{1 +
\frac{\kappa^2}{4\omega_{\rm m}^2}}
\int\limits_0^t d\tau e^{-\gamma_{\rm eff}(t-\tau)}\cdot\nonumber\\
&&\int\limits_0^\tau d\tau'
e^{-\kappa(\tau-\tau')}\left[e^{2i\omega_{\rm m}\tau}\bar{c}^\dag_{\rm in}(\tau')+\bar{c}_{\rm
in}(\tau')\right],\nonumber
\end{eqnarray}
whereas the thermal contribution (\ref{Eq:Solbtherm}) remains unchanged. Evaluating the correlations of the generalized quadrature operators (\ref{Eq:GenQuadrature}) gives the result,
\begin{multline}\label{Eq:XXCorrGen}
\langle\delta X_{\varphi,{\rm I}}(t),\delta X_{\varphi,{\rm I}}(t)\rangle = \\
\frac{G^2}{4\gamma_{\rm eff}\kappa}
\Bigg[N f_- +\frac{1}{2}
+\frac{\kappa^2}{\kappa^2+4\omega_{\rm m}^2}
\left\{N h +\frac{1}{2}\right\}\\
 - {\rm Re}\{M e^{2i\varphi}\} f_+ \Bigg]
+\frac{\gamma_{\rm m}}{\gamma_{\rm eff}}\left(n_{\rm th}+\frac{1}{2}\right),
\end{multline}
where the first two lines describe the contribution from the squeezed environment, whereas the last line contains the thermal contribution.
For clarity, the cumbersome bandwidth dependence is contained in the coefficients $f_\pm$ and $h$,
\begin{align*}
f_{\pm}&=\frac{b_x b_y}{b_y^2\pm b_x^2}\left[\frac{b_y}{b_x+\gamma_{\rm eff}}\pm \frac{b_x}{b_y+\gamma_{\rm eff}}\right],\\
h &= \frac{b_x b_y}{b_y^2- b_x^2}  \left[\frac{b_y\left[b_x^2+b_x\kappa+\frac{\gamma_{\rm eff}}{\kappa}(\kappa^2+4\omega_{\rm m}^2)\right]}{(b_x+\kappa)(b_x^2+2b_x\gamma_{\rm eff}+4\omega_{\rm m}^2)}\right.\\
&\hspace{1.8cm}\left.-\frac{b_x\left[b_y^2+b_y\kappa+\frac{\gamma_{\rm
eff}}{\kappa}(\kappa^2+4\omega_{\rm m}^2)\right]}{(b_y+\kappa)(b_y^2+2b_y\gamma_{\rm
eff}+4\omega_{\rm m}^2)}\right].
\end{align*}
In the white noise limit these functions simplify to
$h,f_\pm \to 1$ and we obtain
\begin{multline}
\langle\delta X_{\varphi,{\rm I}}(t),\delta X_{\varphi,{\rm I}}(t)\rangle =\\
\frac{G^2}{4\gamma_{\rm eff}\kappa}
\left[\left(N+\frac{1}{2}\right)\left(1+\frac{\kappa^2}{\kappa^2+4\omega_{\rm m}^2}\right)-{\rm Re}\{M e^{2i\varphi}\}\right]\\
+ \frac{\gamma_{\rm m}}{\gamma_{\rm eff}}\left(n_{\rm th}+\frac{1}{2}\right)
.\label{Eq:XXCorrWhite}
\end{multline}
Deep in the resolved sideband limit, keeping only terms up to zeroth order in
$\kappa/\omega_{\rm m}$, we recover the simple result (\ref{Eq:XXCorrRS}).


\begin{thebibliography}{99}

\bibitem{Metzger2004}
C. H. Metzger and K. Karrai, Nature \textbf{432}, 1002, (2004).

\bibitem{Gigan2006}
S. Gigan, H. R.B\"ohm, M. Paternostro, F. Blaser, G. Langer, J. B.
Hertzberg, K. C. Schwab, D. B\"auerle, M. Aspelmeyer and A.
Zeilinger, Nature \textbf{444}, 67, (2006).

\bibitem{Arcizet2006b}
O. Arcizet, P. F. Cohadon, T. Briant, M. Pinard and A. Heidmann,
Nature \textbf{444}, 71, (2006).

\bibitem{Schliesser2008}
A. Schliesser, R. Rivi\`{e}re, G. Anetsberger, O. Arcizet and T. J.
Kippenberg, Nature Physics \textbf{4}, 415, (2008).

\bibitem{Corbitt2007}
T. Corbitt, Y. Chen, E. Innerhofer, H. M\"uller-Ebhardt,
 D. Ottaway,
H. Rehbein, D. Sigg, S. Whitcomb, C. Wipf and N. Mavalvala, Phys.
Rev. Lett. \textbf{98} 150802 (2007).

\bibitem{Thompson2008}
J. D. Thompson, B. M. Zwickl, A. M. Jayich, F. Marquardt, S. M.
Girvin and J. G. E. Harris,  Nature \textbf{452}, 72, (2008).

\bibitem{Teufel2008}
J. D. Teufel, J. W. Harlow, C. A. Regal and K. W. Lehnert,
Phys. Rev. Lett. {\bf 101}, 197203, 2008 

\bibitem{Schliesser2009}
A. Schliesser, O. Arcizet, R. Rivi\`{e}re, T. J. Kippenberg,
arXiv:0901.1456v1

\bibitem{Groeblacher2009}
S. Groeblacher, J.B. Hertzberg, M.R. Vanner, S. Gigan, K.C. Schwab, M. Aspelmeyer,
arXiv:0901.1801v1

\bibitem{Walls1983}
D.F. Walls, Nature {\bf 306}, 141 (1983)

\bibitem{Slusher1985}
R.E. Slusher, L.W. Hollberg, B.Yurke, J.C. Mertz, J.F. Valley, Phys. Rev. Lett. {\bf 56}, 788 (1985)

\bibitem{Wu1986}
L.-A. Wu, H.J. Kimble, J.L. Hall, H. Wu, Phys. Rev. Lett. {\bf 57} 2520 (1986)


\bibitem{LaHaye2004}
M. D. LaHaye, O. Buu, B. Camarota, and K. C. Schwab, Science
{\bf 304}, 74 (2004).

\bibitem{Bradaschia1990}
C. Bradaschia et al., Nucl. Instrum. Methods Phys. Res. A
{\bf 289}, 518 (1990); A. Abramovici et al., Science {\bf 256}, 325
(1992); P. Fritschel, Proc. SPIE {\bf  4856}, 282 (2003).

\bibitem{AlmogPRL98}
R. Almog, S. Zaitsev, O. Shtempluck and E. Buks,
Phys. Rev. Lett. \textbf{98} 078103 (2007).

\bibitem{IanPRA78}
H. Ian, Z.R. Gong, Y. Liu, C.P Sun and F. Nori,
Phys. Rev A \textbf{78} 013824 (2008).

\bibitem{ZhangArxiv2009}
J. Zhang, Y. Liu and F. Nori, ArXiv:0902.2526 (2009).

\bibitem{ZhouPRL2006}
X. Zhou and A. Mizel, Phys. Rev. Lett. \textbf{97} 267201 (2006).

\bibitem{RablPRB70}
P. Rabl, A. Shnirman and P.Zoller, Phys. Rev. B \textbf{70}, 205304
(2004).

\bibitem{Mancini2003}
S. Mancini, D. Vitali, P. Tombesi, Phys. Rev. Lett. {\bf 90}, 137901 (2003)

\bibitem{Hammerer2009}
K. Hammerer, M. Aspelmeyer, E.S. Polzik, P. Zoller, Phys. Rev. Lett. {\bf 102}, 020501 (2009)

\bibitem{ClerkNJP10}
A. A. Clerk, F. Marquardt and K. Jacobs, New J. Phys. \textbf{10}
095010 (2008).

\bibitem{VitaliPRA65}
D. Vitali, S. Mancini, L. Ribichini and P. Tombesi, Phys. Rev. A
\textbf{65} 063803 (2002).

\bibitem{TianNJP10}
L. Tian, M.S. Allman and R. W. Simmonds, New J. Phys. \textbf{10}
115001 (2008).

\bibitem{WoolleyPRA78}
M. J. Woolley, A.C. Doherty, G. J. Milburn and K.C. Schwab, Phys.
Rev. A \textbf{78} 062303 (2008).


\bibitem{RuskovPRB71}
R. Ruskov, K. Schwab and A. N. Korotkov, Phys. Rev. B \textbf{71}
235407 (2005).

\bibitem{Wiseman1999}
H.M. Wiseman, J. Opt. B \textbf{1} 459 (1999)

\bibitem{PinardEPL72}
M. Pinard, A. Dantan, D. Vitali, O. Arcizet, T. Briant and A.
Heidmann, Europhys. Lett. \textbf{72} 2005.

\bibitem{ZhangPRA68}
J. Zhang, K. Peng and S. L. Braunstein, Phys. Rev. A \textbf{68}
013808 (2003).

\bibitem{Genes2008}  C. Genes, D. Vitali, P. Tombesi, S. Gigan, and M. Aspelmeyer,
Phys. Rev. A \textbf{77}, 033804  (2008).

\bibitem{Schliesser2006}
A. Schliesser, P. Del'Haye, N. Nooshi, K. J. Vahala, and T. J.
Kippenberg, Phys. Rev. Lett. \textbf{97}, 243905 (2006).

\bibitem{Kippenberg2005}
T. J. Kippenberg, H. Rokhsari, T. Carmon, A. Scherer, and K. J.
Vahala, Phys. Rev. Lett. \textbf{95}, 033901 (2005).

\bibitem{Ritsch1988}
H. Ritsch and P. Zoller, Phys. Rev. A  \textbf{38}, 4657 (1988).

\bibitem{Brownian}
V. Giovannetti and D. Vitali, Phys. Rev. A \textbf{63}, 023812
(2001).

\bibitem{Vahlbruch}
H. Vahlbruch, M. Mehmet, N. Lastzka, B. Hage, S. Chelkowski, A.
Franzen, S. Gossler, K. Danzmann, R. Schnabel, Phys. Rev. Lett.
\textbf{100}, 033602 (2008)

\bibitem{Wilson-Rae2007}
I. Wilson-Rae, N. Nooshi, W. Zwerger, and T. J. Kippenberg Phys.
Rev. Lett. \textbf{99}, 093901 (2007)

\bibitem{Marquardt2007} Florian Marquardt, Joe P. Chen, A. A. Clerk, and S. M. Girvin
Phys. Rev. Lett. \textbf{99}, 093902 (2007)

\bibitem{GenesSim08}
C. Genes, D. Vitali and P. Tombesi, New J. Phys. \textbf{10}, 095009
(2008)

\bibitem{Schliesser2008b}
A. Schliesser, G. Anetsberger, R. Rivière, O. Arcizet, T. J.
Kippenberg, New J. Phys. {\bf 10}, 095015 (2008)

\bibitem{Collett1984}
M. J. Collett and C. W. Gardiner, Phys. Rev. A  \textbf{30}, 1386
(1984).

\bibitem{QuantumNoise} C. W. Gardiner and P. Zoller, {\it Quantum
Noise} (Springer, Berlin, 2000).


\end{thebibliography}
\end{document}